# Automatic code generation from sketches of mobile applications in end-user development using *Deep Learning*

Daniel Baulé[1], Christiane Gresse von Wangenheim[1], Aldo von Wangenheim[1],
Jean C. R. Hauck[1], Edson C. Vargas Júnior[2]

[1]Department of Informatics and Statistics, Federal University of Santa Catarina, Florianópolis, Brazil

[2]Department of Mathematics, Federal University of Santa Catarina, Florianópolis, Brazil

daniel.baule@grad.ufsc.br, c.wangenheimn@ufsc.br, aldo.vw@ufsc.br,
jean.hauck@ufsc.br, edson.junior@ufsc.br

## Abstract

A common need for mobile application development by end-users or in computing education is to transform a sketch of a user interface into wireframe code using App Inventor, a popular block-based programming environment. As this task is challenging and time-consuming, we present the Sketch2aia approach that automates this process. Sketch2aia employs deep learning to detect the most frequent user interface components and their position on a hand-drawn sketch creating an intermediate representation of the user interface and then automatically generates the App Inventor code of the wireframe. The approach achieves an average user interface component classification accuracy of 87,72% and results of a preliminary user evaluation indicate that it generates wireframes that closely mirror the sketches in terms of visual similarity. The approach has been implemented as a web tool and can be used to support the end-user development of mobile applications effectively and efficiently as well as the teaching of user interface design in K-12.

## 1   Introduction

With millions of mobile applications ranging from social media, digital entertainment to medical solutions, a successful and valuable app needs to provide an effective, efficient and pleasurable experience with an engaging user interface (UI). Consequently, user interface design is not just a minor aspect of the development of the app, but essential to the product strategy emphasizing its importance in the mobile app development process (Wassermann 2010).

Typically, the development of UIs involves iterative prototyping, starting with the creative process of making sketches, that are simple hand-drawn representations. They are an effective visual medium to transmit and discuss ideas and compare different alternatives in a simple, quick, and inexpensive way (Huang et al. 2019). From sketches, wireframes that define the visual hierarchy are created, representing the interface layout and structure without visual design details such as colors, images, etc. (Robinson 2019). Once the wireframe is created and revised, it is enhanced by the visual design, until it becomes a high-fidelity prototype (Robinson 2019). When the design is finalized, it is implemented, resulting in the product. As an iterative prototyping process, these steps can be repeated several times during the design of an interface, requiring rework whenever changes are made. In this context, specifically, the process of the generation of code from UI prototypes has a high potential for automation, as it is an uninspired, time-consuming, and error-prone task (Moran et al. 2018; Silva da Silva et al. 2011; Suleri et al. 2019).

While designers typically use graphic editors such as Photoshop or Illustrator to design UIs, a large variety of design tools (e.g., Sketch, Figma, Marvel or Adobe XD) have evolved to become all-in-one tools from designing, prototyping to testing, but still requiring the coding of the UI design. On the other hand, modern IDEs such as Eclipse, Visual Studio, or Android Studio, have powerful interactive drag-and-drop based builders for UI code. But, even with the currently available tools, the transition from sketches or wireframes to code still consists of manually recreating user interfaces (Beltramelli 2019). This also applies to App Inventor (MIT 2020), a popular visual block-based programming environment that has empowered more than 50K users worldwide to create more than 34K mobile applications, with over 30K published on Google Play (Wolber et al. 2015). As such App Inventor represents an important programming environment as more and more mobile applications are being written not by professional software developers but by people with expertise in other domains as well as young people as part of computing education in K-12.

Therefore, solutions are being created for the automatic generation of code for UIs of websites and mobile applications typically adopting Machine Learning (ML) approaches (Baulé et al. 2020). These tools automatically convert hand-drawn sketches or wireframes into front-end code or code representations. This automation facilitates the UI development process saving effort and time as well as helps to prevent accidental mistakes (Ozkava 2019). And, although there exist already several approaches that automatically generate code of mobile applications based on hand-drawn sketches, so far there does not exist a solution that creates App Inventor code (Baulé et al. 2020).

Therefore, we present Sketch2aia, a solution for automatically creating App Inventor code of wireframes based on hand-drawn sketches of mobile application user interfaces. In order to automatically detect UI elements in sketches, we adopt a deep learning approach creating an intermediate representation of the UI that is used to automatically create App Inventor code. The tool is available online and can be used to support the teaching of UI design as part of computing education as well as in end-user development, reducing time spent on the manual creation of wireframes, while at the same time maintaining the creative process of sketching.

## 2 Current Challenges and Proposed Solutions

Considering the importance of the aesthetics and usability of UIs and, thus, the importance of the interface design as well as the potential effort reduction by the automatic generation of UI code, currently there exist only a small number of research efforts in this direction (Baulé et al. 2020) (Table 1). While on the other hand, the relevance of the topic is pointed out by the availability of commercial solutions, such as Microsoft AI Lab's Sketch2Code.

**Table 1** Related approaches

| Reference | Description |
| --- | --- |
| (Aşıroğlu et al. 2019) | An approach for the automation of the code generation process from hand-drawn mock-ups using computer vision techniques and deep learning methods. |
| (Bajammal et al. 2018) | A tool to automate the generation of reusable web components from a mock-up employing visual analysis and unsupervised learning of visual cues to create reusable web components. |
| (Beltramelli 2019) | An approach that uses computer vision and ML to automatically transform wireframe images to high-fidelity mock-ups that can be exported to front-end code such as HTML/CSS. |
| (Beltramelli 2018) | An approach based on a Convolutional Neural Network (CNN) and a Recurrent Neural Network (RNN) for the generation of computer tokens from a single UI screenshot. |

| (Chen et al. 2018) | A deep learning architecture combining CNN and RNN models that distill the crowd-scale knowledge of UI designs and implementations from existing apps to automatically generate a UI skeleton given a UI image. |
|---|---|
| (Chen et al. 2019) | An automated cross-platform UI code generation framework using image processing and deep learning classification techniques to transfer the UI code implementation between two mobile platforms. |
| (Ge 2019) | An approach that searches for visually similar apps using sketches using deep learning to translate sketches into UI structures. Similar UIs are identified by computing a similarity score between structural UI data with the ones in the app repositories. |
| (Halbe and Joshi 2015) | An approach that automatically creates an HTML page from a hand-drawn paper sketch by detecting various HTML controls using ML. |
| (Han et al. 2018) | A method based on object detection and attention mechanisms to automatically generate a web page with CSS style information. |
| (Huang et al. 2018) | A method to automate the transformation of a mockup into a web page by extracting the elements based on the color features of the edges. A bottom-up tag generating method based on Random Forest is proposed to select the tags for elements. The web page is generated by defining HTML/CSS code. |
| (Jain et al. 2019) | An approach that employs a deep neural network (DNN) to detect UI elements in sketches. The output is a platform-independent UI representation object used by a UI parser, which creates code for different platforms. |
| (Kim et al. 2018) | The approach generates HTML code automatically by recognizing web layout based on hand-drawn sketches using computer vision algorithms and Faster R-NN, |
| (Kumar 2018) | An approach based on Beltramelli (2018) and Wallner (2018) to generate HTML code based on hand-drawn website sketches. |
| (Liu et al. 2018) | A framework to transform a UI screenshot into code based on deep learning using a CNN and long short-term memory (LSTM). The model is optimized by a Bidirectional LSTM. |
| (Moran et al. 2018) | An approach that first detects logical components of a UI from a mock-up artifact using computer vision or mock-up metadata. Then, by software repository mining, automated dynamic analysis, CNNs are used to classify UI components into domain-specific types. A data-driven, K-nearest-neighbors algorithm generates a suitable hierarchical UI structure from which a prototype application can be automatically assembled. |
| (Pandian et al. 2020) | An approach that uses deep learning and gestalt laws-based algorithms to convert UI screens into editable blueprints by identifying the UI element categories, their location, dimension, text content, and layout hierarchy. |
| (Robinson 2019) | An approach that uses an artificial neural network (ANN) to translate a wireframe into a normalized image. |
| (Suleri et al. 2019) (Pandian and Suleri 2020) | A prototyping workbench that automatically generates code based on sketches. It generates medium and high fidelity prototypes using UI element detection (MetaMorph) created with a DNN model that detects constituent UI element categories and their position and code generation. With this information, the respective UI elements are created as a medium-fidelity prototype from which executable code is generated. |
| (Wallner 2018) | An approach using ML to code a basic HTML and CSS website based on an image of a design mockup. |
| (Yun et al. 2018) | An approach adopting object detection based on a DNN that detects UI elements by the integration of localization and classification. |

There exist solutions for web UIs as well as mobile applications, focusing more on Android than iOS applications. Yet, so far, no solution specifically for App Inventor apps and or computing education is currently available. Most approaches are based on sketches as input, recognizing the importance of the creative process of sketching as a first step in prototyping the UI in contrast to approaches that aim at complete automation of the UI design process. However, a shortcoming of several approaches is their rather explorative way of focusing on only a very small number of UI components. For example, Aşıroğlu et al. (2019) and Robinson (2019) only consider four different types of components (e.g., Title, Image, Button, Input e Paragraph) and Ge (2019) only seven, limiting their applicability in practice.

Research concerning images of UIs is further complicated due to the unavailability of large datasets that also include sketches. For example, the popular RICO dataset (Deka et al. 2017) being one of the largest datasets on user interfaces of mobile apps with information on over 9.3k Android apps, focuses on the representation of visual, textual, structural, and interactive properties based on screenshots as well as metadata, does not provide, for example, sketches. Thus, considerable effort needs to be spent on the creation of specific datasets. Adopting diverse approaches, some studies captured UI screenshots by crawling the web or app stores. Others generated UI images or sketches synthetically, which may limit the image quality as well as authenticity (Robinson 2019). Yet, this issue and its potential impact on performance and validity that is not further discussed in most cases. An exception is Robinson (2019), who reports a reduction of the performance of the deep learning approach when applied to real sketches, emphasizing, thus, the importance of a great variety in sketches to allow the deep learning approach to better generalize to unseen sketching styles.

For object detection, a large variety of solutions have been created ranging from classic Machine Learning approaches to recent convolutional neural networks (CNNs), and diverse combinations of them. In order to be able to detect context-related information, such as nested UI widgets, some approaches also employ sequence- and structure-analyzing recurrent convolutional neural network-based techniques, including Gated Recurrent Units, LSTM, and Bi-LSTM (Aşıroğlu et al. 2019; Beltramelli 2018; Chen et al. 2018; Han et al. 2018).

In terms of evaluation, the large variety of measures used to analyze the performance of the object detection indicates a lack of a clear standard hindering also the comparison of the approaches. Furthermore, most studies focus (in some cases exclusively) on the analysis of accuracy, not necessarily being aligned with commonly proposed measures for object detection (Zou et al. 2019), such as mAP that has been used only in one research (Suleri et al. 2019). Some studies also evaluate the similarity of the code generated, adopting diverse measures derived from text analysis. Very few also analyze the applicability of the proposed approaches through user studies (Beltramelli 2019; Chen et al. 2018; Robinson 2019; Suleri et al. 2019), yet, so far no empirical study on the application of these approaches in practice has been reported.

## 3 Research Methodology

Based on the results of related work on different kinds of user interfaces, we developed a web tool for the automatic generation of code for wireframes in App Inventor based on sketches following a multi-method approach. Based on the results of a systematic literature review (Baulé et al. 2020), we developed a deep learning model for the detection of UI elements in sketches following the machine learning process proposed by Amershi et al. (2019) and Ashmore et al. (2019):

**Requirements analysis**. During this phase, the main objective of the model and its target features is specified following Mitchell (1997). This also includes the characterization of the inputs and expected outputs, specifying the problem.

**Data management**. During data collection, we searched for available datasets. This includes the selection of available generic datasets for pre-training a model (e.g., ImageNet for image classification tasks), as well as specialized datasets for transfer learning to train the specific model. Due to the unavailability of datasets on user interfaces of App Inventor apps we collected specific data by capturing UI screenshots of App Inventor projects and drawing the respective sketches. The data is prepared by validating, cleaning, and conditioning the data (e.g., to remove duplicates, correct errors, deal with missing values, normalization, data type conversions, etc.). Adopting supervised learning, labels identifying UI elements in sketches were manually assigned using the tool labelme (Wada 2016). The dataset is split into a training set to train the model and a test set to perform an unbiased performance evaluation of the chosen model on unseen data.

**Model learning**. During the model learning phase, an appropriate deep learning framework and machine learning model has been chosen that has been proven effective for a comparable problem or domain, and the volume and structure of the data. Adopting a transfer learning approach, we take a pre-trained network, which was trained on a generic dataset (ImageNet), adapt its output layers to work with our evaluation metrics, and train the input layer and the final, fully connected output layers on our data, initially maintaining the pre-trained internal feature representation structure of the network. This is performed until the network does not further improve. After transfer learning, we unfreeze the internal features of the transfer-trained model allowing all its layers to learn. It then undergoes a training phase called fine-tuning, with the same dataset of the specific domain, where all internal features are finely adjusted to our data. A set of hyperparameters (HYPO) (especially learning momentum and learning rates) for a learning algorithm is selected and dynamically optimized during the fine-tuning process. Here, we trained several similar model variants, with different HYPO, and compared their results.

**Model evaluation**. We defined the metric used for measuring success aligned with the goals to be achieved and the kind of problem faced by the model following commonly used performance measures for object detection (Zou et al., 2019). Then, we tested the model(s) against previously unseen data (test set).

**Wireframe code generation and tool implementation.** As a result, an intermediate representation of the user interface components and their layout contained in the sketches is created. Based on this representation, the respective App Inventor code of the wireframe representation is generated. Therefore, we developed a software module following an iterative and incremental software development approach (Larman and Basili 2003). The results have been integrated into a web application. This included the analysis of the context and requirements, modeling, implementation, and testing. After this initial implementation, integration tests were performed, and based on the results of these tests, necessary corrections and improvements were implemented and tested.

**Preliminary evaluation.** In order to evaluate the quality of the approach, we performed a preliminary evaluation using the Goal Question Metric (GQM) approach (Basili et al. 1994) to define the purpose of the evaluation and to systematically decompose it into quality characteristics. In accordance with the defined characteristics, we collected data by conducting a user study in order to obtain data on the perceived visual similarity of the generated wireframe and the sketch as well as the perceived usefulness and usability of the approach. The collected data was analyzed using descriptive statistics. The results were interpreted taking also into consideration further observations made during the user study.

## 3 Sketch2aia

The purpose of this approach is to allow the automatic generation of wireframes in App Inventor from sketches. Based on the uploaded images of sketches of the UIs of a mobile application, the UI components and layout in the sketches are detected, resulting in an intermediate representation of the UI. Based on this representation, the App

Inventor code of the wireframe is automatically generated and can be downloaded by the user in order to be imported into the App Inventor environment and further developed into a mobile application (Figure 1).

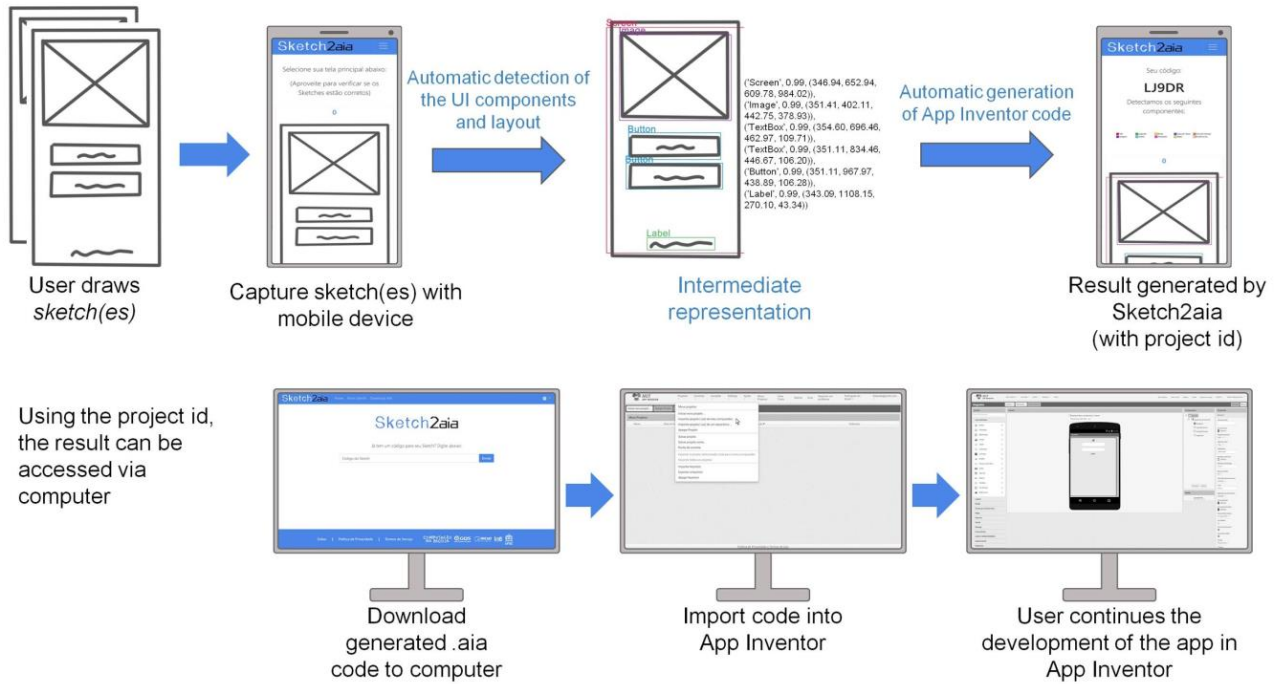

**Fig. 1** Workflow of Sketch2aia

**3.1 Detection of UI Components in Sketches**

This step aims to develop a computer program that learns from an experience (a set of sketches of a mobile app) in relation to the task of detecting UI components and their positions in sketches and measures whether the task performance improves according to experience. Due to the large number of types of UI components, we focused on the most frequently used in App Inventor applications (Alves et al. 2020): Label, Button, Image, TextBox, CheckBox, ListPicker, Slider, Switch, Map.

**Data Preparation.** We created a new data set with hand-drawn sketches of App Inventor applications with the identification of UI components and their positions. Therefore, we collected 165 screenshots of randomly selected applications from the App Inventor Gallery (MIT 2020) and applications developed by the initiative Computing in School (CnE 2020). The screenshots were captured semi-automatically on smartphones. We only selected screens without ads and/or ethical issues. Based on these screenshots, 279 sketches were drawn by 28 volunteers of the initiative Computing in School at the Federal University of Santa Catarina from different fields, educational stages, and ages, following basic instructions. The sketches were manually labeled identifying the UI components and their position using the labelme tool (Wada 2016), resulting in an intermediate representation of the UI design (Figure 2).

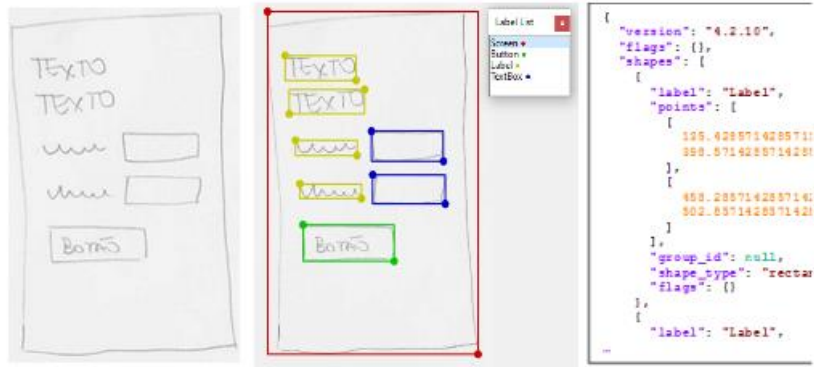

**Fig. 2** Labeling of sketches

Due to the typical usage patterns of UI components in apps, there are considerable differences in the frequency of the different types of components, with Buttons and Labels representing 69% of the components of the data set, and, on the other hand, maps and sliders appearing in only 1% of the sketches. We divided the dataset into a training set with 237 images (~ 85%) used for learning and a test set with 42 images (~ 15%). The dataset is available online: https://bit.ly/Sketch2AIA.

**Model Learning.** We used Darknet (Redmon 2016), an open-source framework for convolutional neural networks to implement the YoloV3 model (Redmon and Farhadi 2019) for detecting objects in images. We adopted a transfer learning strategy from a pre-trained YOLO network with the ImageNet dataset (Redmon and Farhadi 2019) that allows us to use the relatively small dataset to train the network effectively. The fine-tuning of the network was performed following the guidelines of Bochkovskiy (2019), with changes in the convolutional and YOLO layers to suit the number of classes of objects considered (9 types of UI components, in addition to the screen as a whole). The YOLO layers were customized for 10 classes, while the convolutional layers immediately before were configured with 45 filters following Bochkovskiy (2019):

$$number\ of\ filters = (classes\ +\ 5)\ \times\ 3 = (10\ +\ 5)\ \times\ 3 = 45 \quad (1)$$

Following Bochkovskiy (2019), the training set was subdivided into batches of 64 images, and the training was carried out with up to 20,000 batches:

$$max\ batches = classes\ \times\ 2000 = 10\ \times\ 2000\ = 20000 \quad (2)$$

The model is available online https://codigos.ufsc.br/lapix/Sketch2AIA.

**Performance evaluation.** The quality of an object detection model is typically based on the IoU (Intersection Over Union), a measure based on the Jaccard index that assesses the overlap between two bounding boxes, the true bounding box, and the predicted bounding box. Another popular metric for measuring the accuracy of object detectors is Average Precision (AP) (Liu et al. 2018)(Zou et al. 2019). AP is defined as the average detection precision under different recalls and is usually evaluated in an object class-specific manner. To compare performance considering all object classes, the mean Average Precision (mAP) is commonly used as the final metric of performance. In addition to the IoU and mAP, the F1 score is also used to evaluate this type of model. The F1 score is the harmonic average of precision and recall, transmitting information about the balance between these (Robinson 2019). A high F1 score indicates that the system performs a detection that is both accurate and complete.

During the creation of the model, all these metrics were analyzed to enable comparisons with related work. Figure 3 shows the performance evolution of the Sketch2aia model during training according to these metrics.

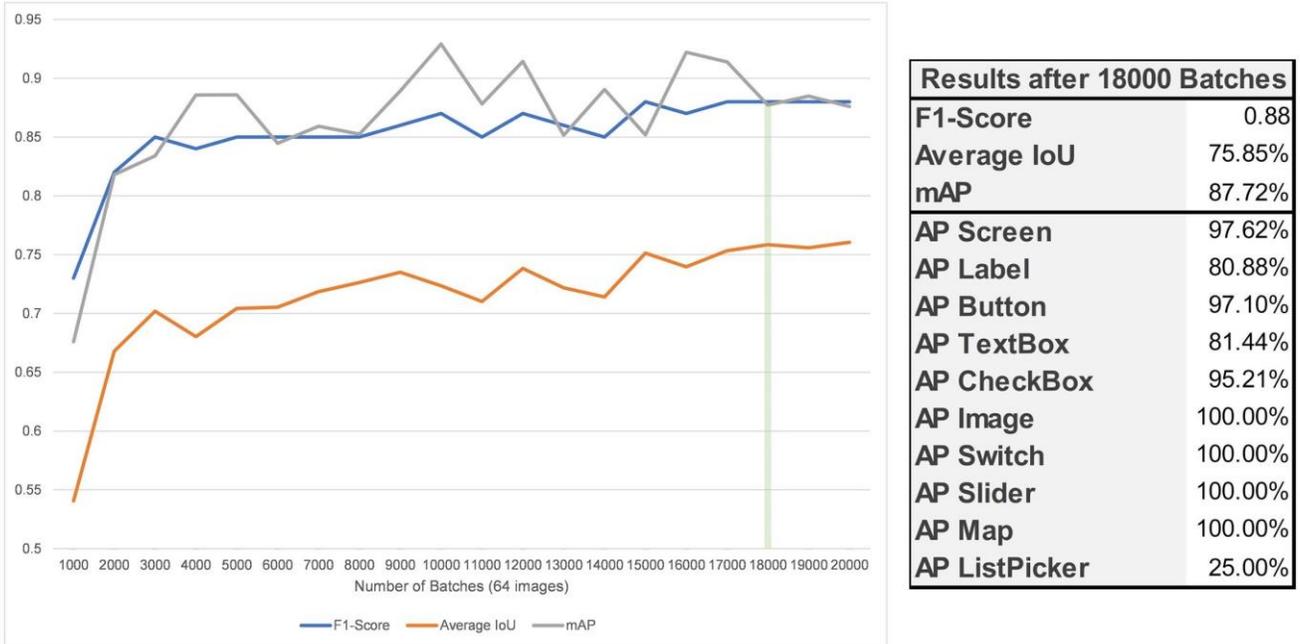

**Fig. 3** Evolution of the performance during training and best results (total and per component)

The best result was obtained after 18,000 batches, presenting an F1 score of 0.88 with more balanced values of mean IoU and mAP, when compared to the other results. The mean Average Precision obtained is 87.72% (Figure 3). Individual accuracy is higher for all types of UI components, except one (ListPicker). Results with 100% accuracy are probably due to the limited size of the test set. Only the component ListPicker has been detected with low accuracy, probably due to its low frequency of appearance in the data set and its similarity with other more frequent components, such as Button and TextBox.

Compared to other models, the average F1 score obtained by Sketch2aia (0.88) is higher than the best score obtained by Robinson (2019) (0.81 for images) and the average score obtained by Huang et al. (2019) (0.775). The mAP of 87.72% obtained by Sketch2aia is also higher than the one reported by Suleri et al. (2019) and Pandian et al. (2020), obtaining a mAP of 84.9%, the only other work reporting this metric. Thus, the Sketch2aia model demonstrates performance equivalent or better than existing models, showing that it supports the detection of UI components with precision.

**3.2 Generating Wireframe Code**

The output of the detection of UI components in sketches is an intermediate representation, in the form of a list of detected objects, the respective confidence levels of classification, and their positions in the image (coordinates of the central point, width, and height) (Figure 4).

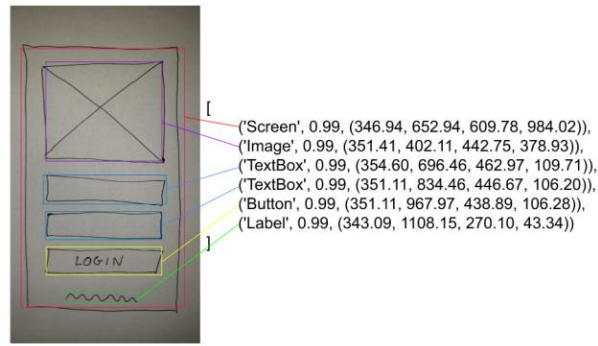

**Fig. 4** Example of the output from the UI component detection in sketches

The first step to generate the App Inventor wireframe code is to eliminate possible overlapping of components. Therefore, each pair of elements is checked if their area of overlap is greater than 50% of the area of the smallest element between them, eliminating the ones with a lower confidence index. Another issue is the definition of the components by their defined positions in space, rather than in the form of a layout as used in App Inventor in order to achieve a more complex organization. Therefore, the components are recursively organized in vertical and horizontal layouts due to their simplicity and more predictable behavior (Figure 5).

```
Function align(components, orientation)
begin
    // Create an empy list to keep components of the current alignment
    currentAllignment = list()
    // Sort list components according to the current orientation
    components = sortList(components, orientation)
    // While component list is not empty
    for each curComponent in components
    begin
        components.remove(curComponent)
        // Check if components in the current aligned components list are aligned with the current one in the opossite orientation
        componentsToAnalize ← list()
        newComponents ← list()
        componentsToAnalize.add(curComponent)
        // For each new aligned component, verify if there are other components aligned to it
        for each analizedComponent in componentsToAnalize
        inicio
            componentsToAnalize.remove(analizedComponent)
            for each component in components
            begin
                // If components are aligned and have not been verified, add and verify
                if analizedComponent.isAligned(component, orientation) then
                    if component not in newComponents then
                        componentsToAnalize.add(component)
                        newComponents.add(component)
                    end_if
                end_if
            end_for
        end_for
        // If not aligned with other components, add to list, otherwise align those components first
        if newComponents.empty() then
            currentAllignment.add(curComponent)
        else
            components ← components - newComponents
            newComponents.add(curComponent)
            currentAllignment.adiciona(align(newComponents, ~orientation))
        end_if
    end_for
    return tuple(orientation, curAlignment)
end
```

**Fig. 5** Algorithm for grouping the detected elements in App Inventor compatible layouts

With the components organized in compatible layouts, a direct translation to a .scm file is performed for each sketch. The App Inventor .scm file encapsulates a JSON structure that contains all the visual components used in the corresponding wireframe, set with empirically determined values, defined based on the analysis of several App Inventor projects. An empty .bky file is also generated for each screen. The .bky file encapsulates an XML structure with all the programming blocks used in the application logic. After generating the wireframe code corresponding to all screens, the properties file is generated according to the project settings (Name, Main Screen, etc.). The generated files are compressed into a zip file following App Inventor's file structure and saved as a .aia file that can be directly imported into App Inventor (Figure 6).

```
{
    "authURL": [
        "ai2.appinventor.mit.edu"
    ],
    "YaVersion": "206",
    "Source": "Form",
    "Properties": {
        "$Name": "Screen1",
        "$Type": "Form",
        "$Version": "27",
        "AppName": "Test",
        "Title": "Screen1",
        "Uuid": "0",
        "$Components": [
            {
                "$Name": "OrganizacaoVertical1",
                "$Type": "VerticalArrangement",
                "$Version": "3",
                "AlignHorizontal": "3",
                "AlignVertical": "1",
                "Height": "-2",
                "Width": "-2",
                "Uuid": "6204202716",
                "$Components": [
                    {
                        "$Name": "ListaSuspensa1",
                        "$Type": "Spinner",
                        "$Version": "1",
                        "AlignHorizontal": "1",
                        "AlignVertical": "1",
                        "Height": "-1",
                        "Width": "-1",
                        "Uuid": "7956601802"
                    },
```

**Fig. 6** App Inventor wireframe code example

### 3.3 Webtool

The Sketch2aia tool was implemented as a web application. The tool allows the user to upload up to 6 images of UI sketches per application (Figure 7). Sketch2aia then automatically detects the UI components and generates the corresponding App Inventor code. The .aia file can then be imported by the user into the App Inventor environment, where s/he can continue developing the application. The tool also allows the user to configure some parameters, such as the project name, the main screen and choosing using PickList or SuspendedList, two components with a similar function in App Inventor, but being aesthetically different.

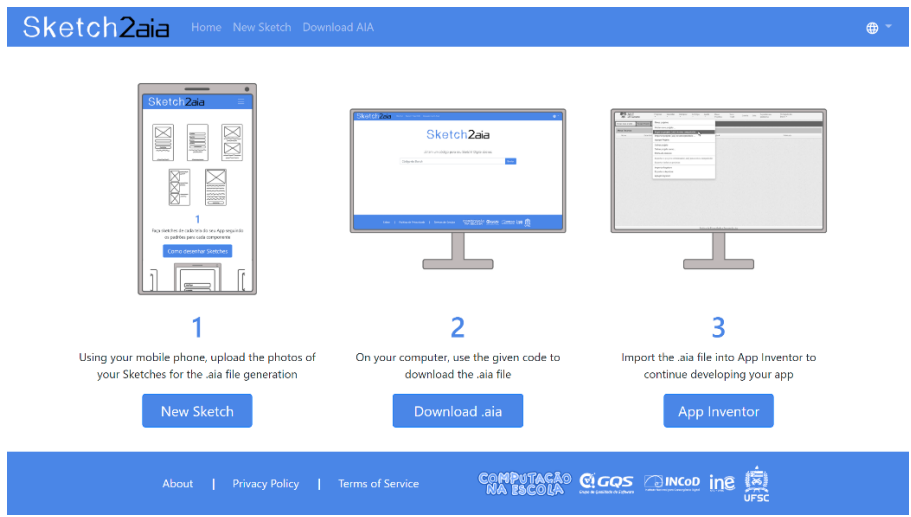

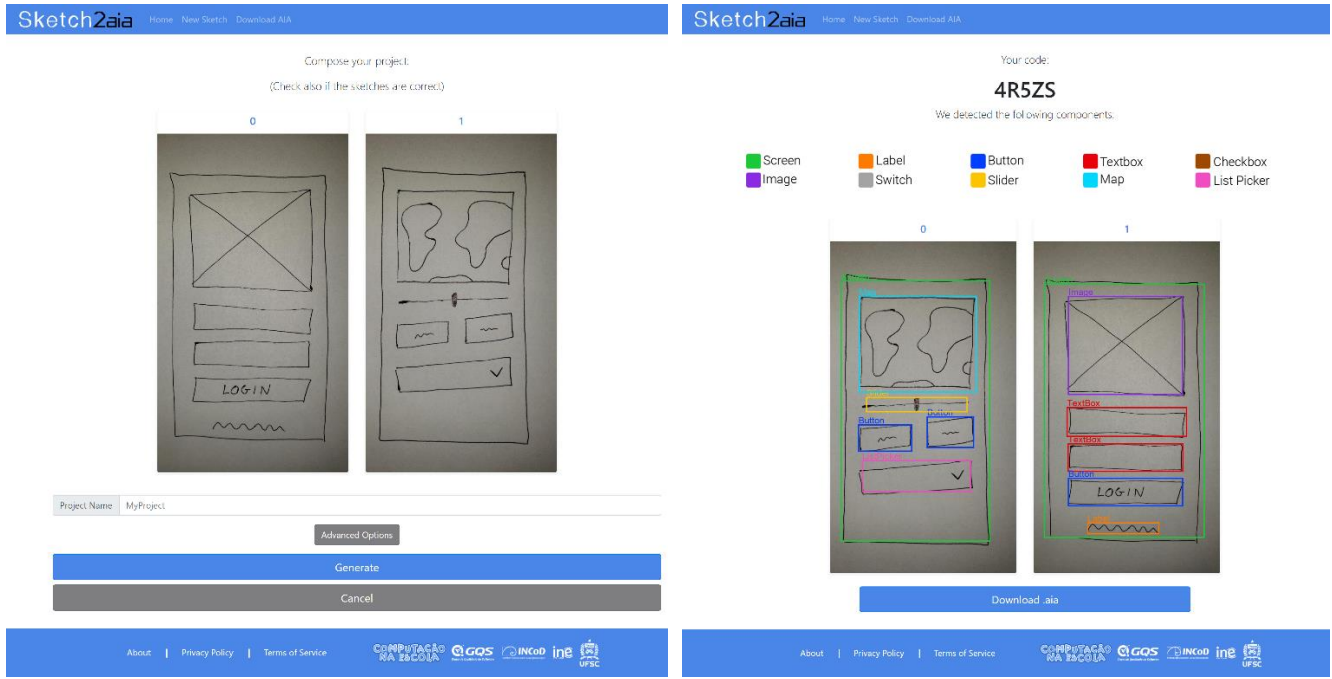

**Fig. 7** Example screenshots of the web tool

It is possible to preview the result of the object detection, enabling the identification of differences between the expected components and those actually detected, before importing the code into the App Inventor environment. The code generated by the tool also makes it easy to share projects and download them onto a computer, even if the images of the sketches were uploaded via smartphone.

The web tool is available online: http://apps.computacaonaescola.ufsc.br/s2a.

## 4. Evaluation

We conducted a user study to evaluate whether the generated wireframes correspond to the user interface design of the sketches as well as the usefulness and usability of the tool. In this study, we presented 10 pairs of randomly selected sketches and screenshots of wireframes generated with Sketch2aia to the participants. In addition, participants were also asked to draw sketches of a mobile application and use the Sketch2aia tool for generating their wireframe code. They were then asked to download the .aia code created and to import it into App Inventor, running the application's user interface on their smartphones. The participants assessed the correspondence of each of the user interface designs of the wireframes with the sketches on an ordinal 4-point scale, ranging from not corresponding to totally corresponding. As part of the study, we also evaluated the perceived usefulness (through one question on a nominal scale) and usability of the sketch2aia approach by adopting the System Usability Scale (SUS) (Brooke 1986), asking the participants to evaluate the usefulness and usability after they created the wireframe as part of the evaluation. We also asked the participants to indicate strengths and weaknesses in order to identify improvement opportunities.

The user study was conducted with members of the Computing at School initiative at the Federal University of Santa Catarina/Brazil in June 2020. Twenty-two students, researchers, and professors with a background in computer science and/or design participated, as well as 6 young people and 1 teacher representing users in a K-12 educational context. The study was conducted online, providing detailed instructions to users, along with a questionnaire. Participation was voluntary, with a response rate of 75.68%.

According to the responses of the participants, the majority of the generated wireframes correspond largely or totally to the sketches of the 10 examples we provided, with similar results to the sketches drawn by the participants themselves, as shown in Figure 8.

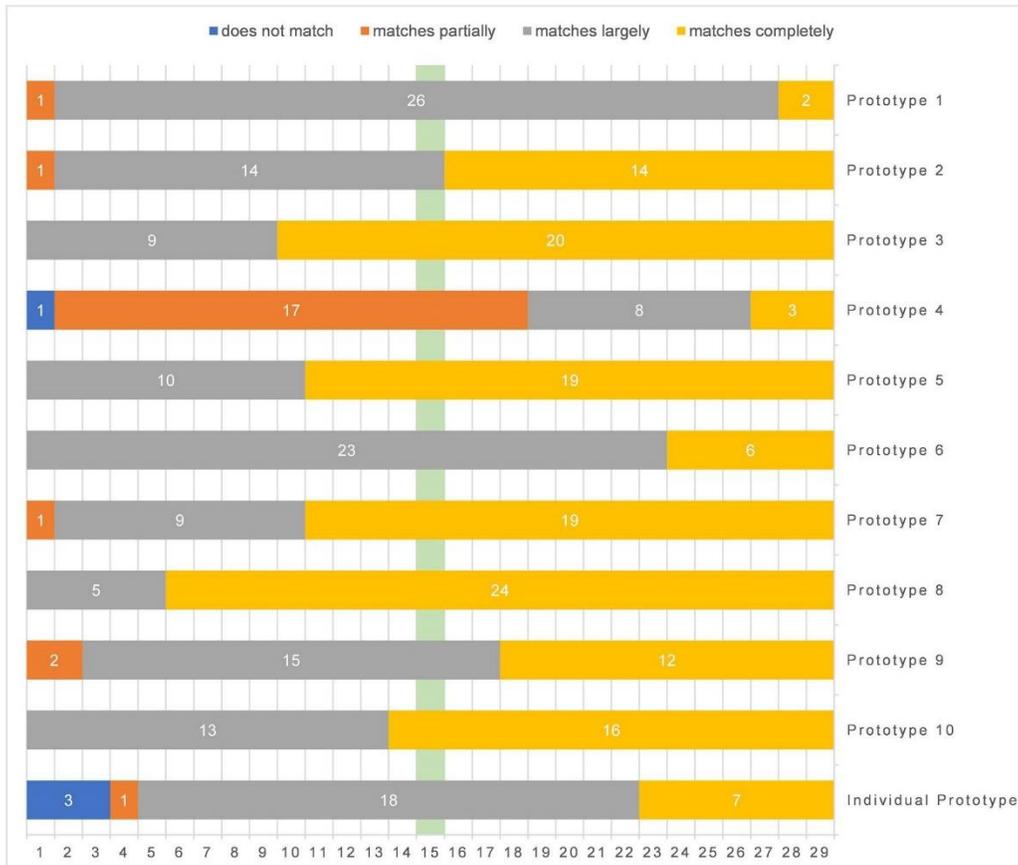

**Fig. 8** User evaluation results and examples of generated wireframes

An exception is prototype 4, which corresponds little according to most participants (Figure 9). This result is probably due to the incorrect detection of the CheckBox at the top of the screen, due to the drawing style used for the Label in the sketch with a combination of letters and placeholders. Most of the issues indicated in the user study refer to the layout of the generated UI, such as the alignment of the UI elements (Figure 9). We also identified an issue regarding the vertical distribution of the UI components, which might be resolved by including white spaces in the App Inventor wireframe for a better approximation with the sketch layout.

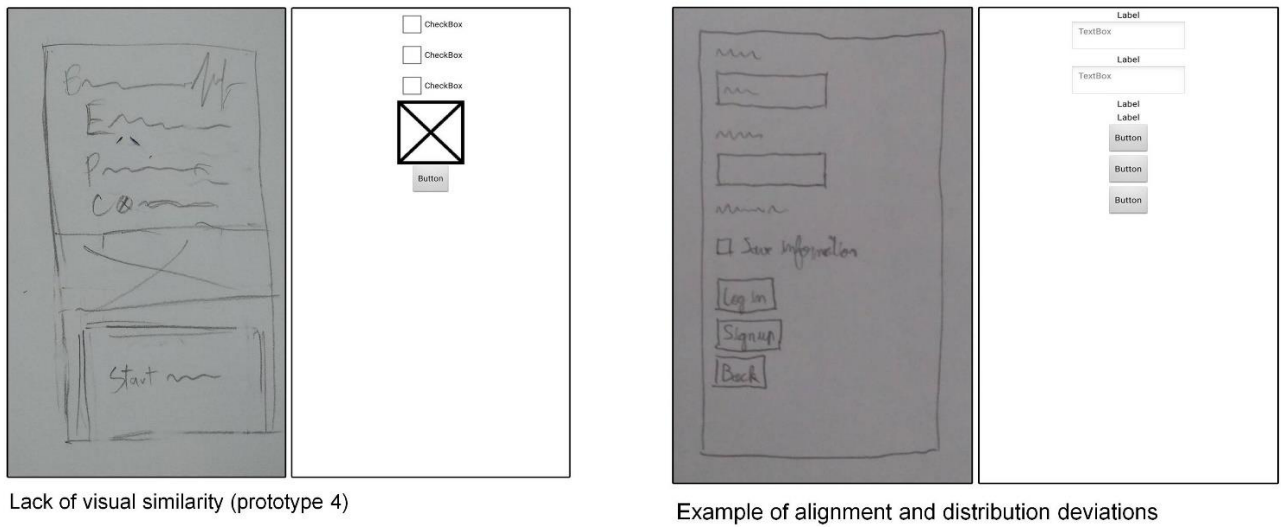

**Fig. 9** Examples of sketches and generated wireframes

Positive results were also obtained concerning the tool's applicability and usability. 100% of the participants considered the Sketch2aia tool to be useful in the app development process and perceived the usability as excellent with an average *SUS* score of 92.16 following the acceptability ranges as defined by Bangor et al. (2008).

The main strengths of the web tool highlighted are its practicality, ease of use, and time savings in the app development process with App Inventor. On the other hand, the main weakness mentioned by the participants is related to the organization, size, and layout of the elements positioned in the wireframes, as well as the misclassification of certain elements that may be drawn similar in sketches, such as labels and switches. Some suggestions were beyond the scope of our current research, such as supporting also the generation of web interfaces, textual recognition for the automatic generation of the text of user interface elements as well as support for the next steps in the user interface design, including the selection of a color palette as part of the visual design.

## 5. Discussion

These results provide a first indication that the Sketch2aia approach can accurately detect objects in sketches of mobile applications as well as automatically generate App Inventor code of visually similar wireframes. Considering that the performance evaluation of object detection achieved an F1 score of 0.88, indicates that the approach can identify the UI components in sketches correctly. Except for ListPicker, all types of components achieved an AP above 80%. The performance below-average observed with respect to the element ListPicker is probably due to its low frequency of appearance in the data set and its similarity with other more frequent components, such as Button and TextBox. In the future, the introduction of more examples of ListPicker in the data set, or even the creation of a different pattern with more distinct characteristics for its representation in sketches could help to solve these problems. Comparing these results with the state of the art, the quality of the approach becomes also evident, as it demonstrates performance values above the ones reported by related works that evaluated the accuracy/precision of their approaches (Baulé et al. 2020) (Table 2).

**Table 2** Comparison with evaluation results of related work

| Reference | Main findings | | |
|---|---|---|---|
| | **Accuracy/Precision and similarity measures** | **Processing time** | **Usability** |
| (Aşıroğlu et al. 2019) | The model achieves 96% method accuracy and 73% validation accuracy. | -- | -- |
| (Bajammal et al. 2018) | VizMod achieves on average 94% precision and 75% recall in terms of agreement with the developers' assessment. | -- | -- |
| (Beltramelli 2019) | -- | 24 times speed increase on average in interface design taking the tool only 170 sec to create high-fidelity prototype and front-end code. | -- |
| (Beltramelli 2018) | Pix2code can automatically generate code from a single input image with over 77% accuracy. Minimum classification error ranges from 11.01% (web) to 22.34% (Android) and 22.73% (iOS). | -- | -- |
| (Chen et al. 2018) | Accuracy: 60.28% of the generated UI skeletons exactly match the ground truth UI skeletons. Accuracy degrades when the UI skeletons are too simple (≤ 10 UI components, ≤ 3 containers, and/or ≤ 3 depth). Similarity ratings for the experiment and control group are 4.2 vs. 3.65. | The experimental group implements the skeleton UIs faster than the control group implementing the design from scratch (with an average of 6.14 min vs. 15.19 min). | On average, satisfactory ratings for the experimental and control group are 4.9 versus 3.8. |
| (Chen et al. 2019a) | The CNN classification achieves more than 85% accuracy. | -- | -- |
| (Halbe and Joshi 2015) | Preliminary empirical evaluation of the accuracy of results obtained of at least 70%. | -- | -- |
| (Huang et al. 2018) | Precision varies from 0.419 (FORM) to 0.911 (SPAN leaves), Recall varies from 0.499 (P) to 0.952 (SPAN leaves), F1 value varies from 0.128 (SPAN inner nodes]) to 0.931 (SPAN leaves). Accuracy varies from 0.651 (DIV) to 0.964 (SPAN inner nodes). | -- | -- |
| (Liu et al. 2018b) | 85% accuracy on the test set. | -- | -- |
| (Moran et al. 2018) | The precision of the CNN approach is 91.1%. | -- | REDRAW has shown potential for industrial design and development workflows, yet requires to be adjusted to specific processes. |
| (Robinson 2019) | Precision varies from 0.562 (paragraph) to 0.896 (image), Recall varies from 0.461 (paragraph) to 0.741 (image), F1 score varies from 0.548 (paragraph) to 0.811 (image). Button and input cause confusion for small elements. Users top rated the deep learning approach with 22/22 votes with respect to similarity. | -- | -- |
| (Suleri et al. 2019) (Pandian and Suleri 2020) | MetaMorph detects UI elements from low fidelity sketches with 84.9% mAP with 72.7% AR (Pandian et al. 2020). | -- | 87% of the participants would like to use the tool frequently to create prototypes. 80% of the users found the tool easy to use. None of the users thought that they would require any technical support to use the system or that the system is unnecessarily complex. The tool scored an average of 78.5 points out of 100, which implies overall good usability. |

The results of the user study also indicate that the wireframes generated by the tool largely correspond to the respective sketches. Most of the issues mentioned by the participants refer to the layout of the generated UI, such as the size and alignment of the user interface elements (Figure 9). As part of these issues related to the organization of the UI elements and the wireframe layouts, we also observed that the component spacing could be improved with the inclusion of blank spaces being a possible solution for a better approximation of the sketch layout.

Performance tests also indicate that the tool can be an efficient way to reduce the UI design effort as it takes Sketch2aia only about 58 seconds to generate wireframes for a three-screen application, especially when considering that a UI/UX developer can take an average of more than 1 hour to create a high-fidelity prototype (Beltramelli 2019).

Different than other approaches, as, e.g., REDRAW (Moran et al. 2018), Sketch2aia can be used to prototype several screens (with a maximum of 6) of an application at the same time, eliminating the need to individually prototype each screen and, then manually combine them in a single application. Sketch2aia is also capable of detecting a set of nine user interface components most used in App Inventor applications, overcoming very small sets, which limit some related approaches. For example, Aşıroğlu et al. (2019) only consider four different types of elements such as TextBox, Dropdown, Button, and CheckBox. Similarly, Robinson (2019) also considers only four types of UI elements and (Ge, 2019) only seven.

And, rather than creating code for professional mobile development, the generation of App Inventor code also provides currently inexistent support for end-user development of mobile apps as well the teaching of computing using the popular block-based programming environment App Inventor that allows anyone to create mobile applications.

**Threats to validity.** Due to the characteristics of this type of research, we identified possible threats and applied mitigation strategies to minimize their impact. To create an authentic and varied data set, we manually created sketches for the App Inventor application user interfaces involving 28 participants, in order to reduce the risk that the images do not adequately represent realistic sketches designed by a target audience with different backgrounds and ages. Furthermore, the sample size of 279 UI sketches of App Inventor applications selected at random from the App Inventor Gallery and of applications developed by the initiative Computing at School at the Federal University of Santa Catarina provides a representative sample of App Inventor application user interfaces.

The selection of the deep learning model was systematically based on the literature and results of testing and comparing different versions of YOLO to maximize performance results. Performance measures were selected based on common measures used for this type of application to ensure an adequate assessment.

Regarding the evaluation, we systematically defined the study using the GQM approach (Basili et al. 1994). In comparison to evaluations reported by related works, a comparison of a total of 11 sketches with the generated wireframes is considered acceptable to provide first ideas about the correspondence of the results. Regarding the sample size, our evaluation used data collected from 29 participants. In the context of a preliminary assessment, this is an acceptable sample size that allows the first results to be obtained. However, as the data were collected in only one context, further studies are needed to increase external validity.

# 6 Conclusion

In this article, we introduce Sketch2aia, an approach for automatically prototyping user interfaces of mobile applications with App Inventor. A preliminary evaluation of the approach demonstrates that it is capable of accurately detecting and classifying user interface components and their position in a sketch and generating App Inventor code of wireframes that are visually similar to the sketches. First feedback also indicates that the approach may effectively and efficiently support the user interface design process as part of the development of mobile apps. In the future, we are planning to amplify the evaluation of the approach on a larger scale as well as to enhance the approach including the recommendation of user interface guidelines and examples.

## Acknowledgments

We would like to thank all participants for their help in drawing sketches and providing feedback.

This work was supported by CNPq (*Conselho Nacional de Desenvolvimento Científico e Tecnológico – www.cnpq.br*), an entity of the Brazilian government focused on scientific and technological development.